\documentclass[11pt]{article}
\newtheorem{theorem}{Theorem}
\newtheorem{axiom}[theorem]{Axiom}

\begin{document}

\title{\textbf{Possible Effects of Spacetime Foam in Particle Physics }}
\author{A.A. Kirillov \\
%EndAName
\emph{Institute for Applied Mathematics and Cybernetics,}\\
603005 \emph{Nizhnii Novgorod, Russia}}
\date{}
\maketitle

\begin{abstract}
We present an extension of quantum field theory to the case when spacetime
topology fluctuates (spacetime foam). In this extension the number of
bosonic fields becomes a variable and the ground state is characterized by a
finite particle number density. It is shown that when the number of fields
remains a constant, the standard field theory is restored. However, in the
complete theory the ground state has a nontrivial properties. In particular,
it produces an increase in the level of quantum fluctuations in the field
potentials and an additional renormalization of masses of particles. \ We
examine fluctuations of massless fields and show that in the presence of a
temperature (thermal state) these fluctuations has $1/f$ spectrum. In the
case of electromagnetic field this should produce an additional $1/f$ -\
noise in electric circuits.
\end{abstract}

\section{Introduction}

As was first pointed out by Wheeler at Planck scales the topology of space
experiences quantum fluctuations (the so-called spacetime foam) \cite
{wheeler} and since then, many attempts had been done to include the
spacetime topology in the number of dynamical characteristics, e.g., see
Refs. \cite{wormholes,tch}. The existence of such fluctuations means that
the standard classical picture, in which the spacetime represents a smooth
manifold, has only an approximate character and is apparently valid after
some averaging out. In this sense one may speak only about an effective
manifold corresponding to the physical space. The need to include topology
changes in quantum gravity can be easily illustrated by the analogy with
relativistic particles. Indeed, in particle physics there exists a minimal
uncertainty in measurements of the position of a particle $\Delta x_{\min
}\sim 1/m$ (the Compton length). In the processes, when $\Delta x_{\min }$
may be considered as a small value (in comparison with characteristic scales
involved), the number of particles remains a constant, while in the opposite
case it represents an extra dynamical variable. We note that the number of
particles is the simplest topological characteristic of the system, which
consist of a set of particles, and, therefore, the change of this number
gives the simplest example of topology transformations. In quantum gravity
there also exists a similar uncertainty $\Delta g_{\min }\sim
L_{pl}^{2}/L^{2}$ (where $L_{pl}$ is the Planck length and $L$ is a
characteristic size of the region in which the gravitational field is
measured). Therefore, quantum gravity processes, in which $\Delta g_{\min }$
cannot be considered as a negligible value, have unavoidably to involve
topology changes.

In this paper we will use the axiomatic approach (suggested in Ref. \cite
{k99}) to construct the extension of quantum field theory to the case when
the topology of spacetime is a variable and to examine possible effects of
the spacetime foam in particle physics. This approach is based on a few
basic principles (or axioms) and results in a generalization of the method
of second quantization (we call it as a second quantization of distributed
systems). The first axiom can be formulated as follows:

\begin{axiom}
\textit{All possible topologies of the effective physical space have to be
described in terms of the basic Minkowski space.}
\end{axiom}

This axiom is not arbitrary but comes from the analysis of the role of the
measurement procedure in quantum theory, e.g., see, Ref. \cite{Landau}.
Indeed, in quantum mechanics the measurement device has to obey classical
laws (with a sufficient degree of accuracy) and, therefore, it always
introduces peculiar restrictions on possible ways to describe and classify
quantum states. In particular, spatial relationships are enforced by
measurement instruments and thus, the fundamental role must play a basic
space which comes from the measuring device. In the modern physical picture
the role of such a basic space plays the ordinary Minkowski space (at least
at laboratory scales). On the contrary, the effective manifold, which can be
related to the physical space, plays a secondary role and is not, in
general, directly observable. Moreover, in the case of an arbitrary quantum
state this manifold does not possesses a definite structure at all. We note
that analogous principle was first used in Ref. \cite{log} (see also
references therein).

From this axiom we derive the following important consequence. Since in the
general case the topology of the effective space differs from that of the
basic space, therefore, upon the transformation to the basic space all
physical observables will be described by multi-valued functions of
coordinates. We can say that physical observables may admit a set of images
in the Minkowski space and a change of the topology corresponds to the
respective change in the number of images.

The next axiom represents the so-called identity principle:

\begin{axiom}
\textit{Images of the same physical observable should obey the identity
principle.}
\end{axiom}

This axiom generalizes the principle of indistinguishability of particles to
the case of fields and other possible observables and it results in the fact
that images of an arbitrary observable should obey to a definite statistics
(Bose or Fermi). To fix the choice of the statistics we shall use an
additional axiom as follows:

\begin{axiom}
\textit{The statistics should be unique for all types of observables.}
\end{axiom}

This axiom comes from the fact that the number of images of an arbitrary
observable is determined by the topology of the effective Riemannian space.
In the most general case one can construct a quantum theory in which the
last axiom does not work. However, in this case the relation of the number
of images of observables to the topology of the effective space breaks down.

It follows, from the last two axioms, that \textit{the only acceptable
choice is Fermi -Dirac statistics}, since otherwise we will be unable to
include in this scheme fermions.

Now, using these basic principles we can construct an extension of quantum
field theory which is based on a generalization of the second quantization
scheme. Beforehand, we recall that at a fundamental level, the composition
of matter is determined by a set of fields and their sources. The sources
are point particles, which in the standard quantum theory behave like
fermions. The second quantization of sources gives the standard description
of fermions exactly the same as in particle physics (we recall that by
assumption there is only Fermi statistics). Thus the extension pointed out
above involves bosonic fields.

\section{Second quantization of distributed systems}

Let $M$ be the basic manifold (Minkowski space), $\varphi $ be an arbitrary
field on $M$, and $A$ be a set of possible readings of an elementary
detector (the elementary system of quantum numbers). More detailed
consideration of this section (and subsequent section) can be found in Ref. 
\cite{k99}. Elements of $A$ are a set of numbers $\xi =$ $\left( \eta ,\zeta
\right) $, where $\zeta $ describes the position in space $M^{\ast }$ at
which the elementary measurement takes place (in what follows $M^{\ast }$
denotes either the space $M$ or the mode space) and $\eta \in V$ denotes
either the field amplitude or the number of particles corresponding to the
field. The result of a complete measurement of the field $\varphi $ is the
map $\varphi :M^{\ast }\rightarrow A$, which in the standard theory can be
represented by functions $\eta \left( \zeta \right) $ on $M^{\ast }$ with
values in $V$.

The extension of the quantum field theory appears if we introduce the real
physical space $M_{ph}$, which is assumed to have an arbitrary topology, and
consider physical fields as an extended map of the form $\varphi ^{\ast
}:M_{ph}^{\ast }\rightarrow A$. Thus, the result of a real measurement will
be a set of functions $\left\{ \eta _{J}\left( \zeta \right) \right\} $,
where $J=\left( i,\sigma _{i}\right) $, $\zeta \in \sigma _{i}$, and $\sigma
_{i}\subset M^{\ast }$ is the range of the function $\eta _{J}\left( \zeta
\right) $. Note that in general the dimensionality of the pieces $\sigma
_{j} $ can differ from that of $M^{\ast }$. Formal classification of such
states can be reached as follows.

Let us introduce a set of operators $C^{+}\left( \xi \right) $ and $C\left(
\xi \right) $, the creation and annihilation operators for an individual
element of the set $A$. For the sake of simplicity we assume that
coordinates $\xi $ take discrete values. The Fermi statistics implies that
these operators should satisfy the anticommutation relations 
\begin{equation}
\left\{ C\left( \xi \right) C^{+}\left( \xi ^{\prime }\right) \right\}
_{+}=\delta _{\xi \xi ^{\prime }}.  \label{CCR}
\end{equation}
We define the vacuum state $\left| 0\right\rangle $ by the relationship $%
C\left( \xi \right) \left| 0\right\rangle =0$ and build a Fock space $F$ in
which the basis consists of the vectors ($n=1,2,\ldots $) 
\begin{equation}
\left| \xi _{1},\xi _{2},\ldots ,\xi _{n}\right\rangle
=\prod_{i=1}^{n}C^{+}\left( \xi _{i}\right) \left| 0\right\rangle .
\label{FSTS}
\end{equation}
The vacuum state corresponds to complete absence of a field and hence of all
observables associated with the field. The state $\left| \xi \right\rangle $
describes the field $\varphi $ with only one degree of freedom. This can be
either a field concentrated at a single point or a field containing only one
mode. States described by single-valued functions are constructed as
follows: 
\begin{equation}
\left| \eta \left( \zeta \right) \right\rangle =\prod_{\zeta \in M^{\ast
}}C^{+}\left( \eta \left( \zeta \right) ,\zeta \right) \left| 0\right\rangle
.  \label{SS}
\end{equation}
Let us introduce the creation operator for a complex of degrees of freedom 
\begin{equation}
D^{+}\left( \eta \left( \zeta \right) ,\sigma \right) =\prod_{\zeta \in
\sigma }C^{+}\left( \eta \left( \zeta \right) ,\zeta \right) ,
\end{equation}
where the domain of the function $\eta \left( \zeta \right) $ is limited to
the set $\sigma \in M^{\ast }$. Then the states with an arbitrary number of
fields can be written 
\begin{equation}
\left| \eta _{1},\eta _{2}\cdots ,\eta _{n}\right\rangle
=\prod_{i=1}^{n}D^{+}\left( \eta _{i}\left( \zeta \right) ,\sigma
_{i}\right) \left| 0\right\rangle .  \label{sts}
\end{equation}

\section{Example of a scalar field}

Consider as an example a real scalar field $\varphi $ and for the sake of
convenience we put the field in a box with length $L$ and impose periodic
boundary conditions. Consider the expansion of the field $\varphi $ in plane
waves, 
\begin{equation}
\varphi \left( x\right) =\sum_{k}\left( 2\omega _{k}L^{3}\right)
^{-1/2}\left( a_{k}e^{ikx}+a_{k}^{+}e{\ }^{-ikx}\right) ,  \label{field}
\end{equation}
where $\omega _{k}=\sqrt{k^{2}+m^{2}}$, and $k=2\pi n/L$, with $n=\left(
n_{x},n_{y},n_{z}\right) $. The general expression for the Hamiltonian is 
\begin{equation}
H=H_{0}+V\left( \varphi \right)  \label{Hf}
\end{equation}
where $H_{0}$ describes free particles 
\begin{equation}
H_{0}=\sum_{k}\omega _{k}a_{k}^{+}a_{k}\,\,  \label{H}
\end{equation}
and the potential term $V$ is responsible for the interaction. In the case
of the scalar field the set $A$ represents pairs $\xi =\left( a,k\right) $.
The generalized second quantization results in the fact that the number of
modes for every $k$ becomes a variable and, therefore the set of variables $%
\left\{ a_{k},a_{k}^{+}\right\} $ is replaced by the somewhat expanded set $%
\left\{ a_{k}\left( j\right) ,a_{k}^{+}\left( j\right) \right\} $, where $%
j\in \left[ 1,\ldots N_{k}\right] $, and $N_{k}$ is the number of modes for
a given wave number $k$. For a free field the energy is an additive
quantity, which can be written 
\begin{equation}
H_{0}=\sum_{k}\sum_{j=1}^{N_{k}}\omega _{k}a_{k}^{+}\left( j\right)
a_{k}\left( j\right) \,\,.
\end{equation}
Since the modes are indistinguishable, the interaction operator can be
presented in the form (e.g., see Ref. \cite{k99}) 
\begin{equation}
V=\sum_{n,m,m^{\prime }}\sum_{k_{1},\ldots k_{n}}^{\prime
}\sum_{j_{1},\ldots j_{n},}V_{\left\{ m\right\} ,\left\{ m^{\prime }\right\}
}^{n}\left( k_{1},\ldots k_{n}\right) \prod_{i=1}^{n}\left(
a_{k_{i}}^{+}\left( j_{i}\right) \right) ^{m_{i}}\left( a_{k_{i}}\left(
j_{i}\right) \right) ^{m_{i}^{\prime }},
\end{equation}
where the indices $j_{i}$ run through the corresponding intervals $j_{i}\in %
\left[ 1,\ldots N\left( k_{i}\right) \right] $ and we assume that the sum
with respect to the wave vectors $k_{i}$ contains no terms with equal
indices, i.e., $k_{i}\neq k_{j}$ for any pair of indices $i$ and $j$.

The definition of $V$ contains an ambiguity which corresponds to the
ambiguity in possible ways to generalize terms of the type $\left( a\right)
^{m}$, where $m>1$, e.g., one can use one of the expressions 
\begin{equation}
\left( a_{k}\right) ^{m}\rightarrow \left\{ 
\begin{array}{c}
\sum_{j=1}^{N}\left( a_{k}\left( j\right) \right) ^{m}, \\ 
\sum_{j_{1},j_{2}}\left( a_{k}\left( j_{1}\right) \right) ^{m-1}a_{k}\left(
j_{2}\right) , \\ 
\cdots \\ 
\sum_{j_{1},\ldots j_{m}}a_{k}\left( j_{1}\right) a_{k}\left( j_{2}\right)
\ldots a_{k}\left( j_{m}\right) ,
\end{array}
\right.
\end{equation}
or any their combination with the demand that in the classical theory (when $%
N_{k}=1$ for every $k$) we should obtain the correct classical expressions.
It is remarkable, that this ambiguity reflects the renormalization procedure
in the conventional quantum field theory and may provide a more deep
physical ground for the renormalization theory.

Consider now the representation of occupation numbers. First, we introduce
the notation 
\begin{equation}
A_{m,n}\left( k\right) =\sum_{j=1}^{N\left( k\right) }\left( a_{k}^{+}\left(
j\right) \right) ^{m}\left( a_{k}\left( j\right) \right) ^{n}.  \label{A2}
\end{equation}
Then the expression for the field Hamiltonian takes the form 
\begin{equation}
H=\sum_{k}\omega _{k}A_{1,1}\left( k\right) +\sum_{n,\left\{ m\right\}
,\left\{ m^{\prime }\right\} }\sum_{k_{1},\ldots k_{n}}^{\prime }V_{\left\{
m\right\} ,\left\{ m^{\prime }\right\} }\left( k_{1},\ldots k_{n}\right)
\prod_{i=1}^{n}A_{m_{i},m_{i}^{\prime }}\left( k_{i}\right) .  \label{Ham}
\end{equation}
The fundamental operators $C^{+}\left( \xi \right) $ and $C\left( \xi
\right) $ have the representation 
\begin{equation}
C\left( a^{\ast },k\right) =\sum_{n=0}^{\infty }C\left( n,k\right) \frac{%
\left( a^{\ast }\right) ^{n}}{\sqrt{n!}},\;\;C^{+}\left( a,k\right)
=\sum_{n=0}^{\infty }C^{+}\left( n,k\right) \frac{a^{n}}{\sqrt{n!}},
\label{C}
\end{equation}
where $C\left( n,k\right) $ and $C^{+}\left( n,k\right) $ obey the
anticommutation relations 
\begin{equation}
\left\{ C\left( n,k\right) ,C^{+}\left( m,k^{\prime }\right) \right\}
=\delta _{n,m}\delta _{k,k^{\prime }}\,\,.  \label{3.3m}
\end{equation}
The physical meaning of the operators $C\left( n,k\right) $ and $C^{+}\left(
n,k\right) $ is that they create and annihilate modes with a given number of
particles. In terms of these operators we get

\begin{equation}
\widehat{A}_{m_{1},m_{2}}\left( k\right) =\sum\limits_{n=0}^{\infty }\frac{%
\sqrt{\left( n+m_{1}\right) !\left( n+m_{2}\right) !}}{n!}C^{+}\left(
n+m_{1},k\right) C\left( n+m_{2},k\right)  \label{A}
\end{equation}
and expression for the Hamiltonian in terms of the operators $C^{+}\left(
\xi \right) $ and $C\left( \xi \right) $ can be obtained by substituting (%
\ref{A}) into (\ref{Ham}). For a free field, the eigenvalues of the
Hamiltonian take the form 
\begin{equation}
\widehat{H}_{0}=\sum_{k}\omega _{k}\widehat{A}_{1,1}\left( k\right)
=\sum_{n,k}n\omega _{k}N_{n,k}\,,  \label{zu}
\end{equation}
where $N_{n,k}$ is the number of modes for fixed values of the wave number $%
k $ and the number of particles $n$ ($N_{n,k}=C^{+}\left( n,k\right) C\left(
n,k\right) $).

Thus, the field state vector $\Phi $ is a function of the occupation numbers 
$\Phi \left( N_{k,n},t\right) $, and its evolution is described by the
Schrodinger equation 
\begin{equation}
i\partial _{t}\Phi =H\Phi .
\end{equation}

\section{Physical particles and effective field}

Consider now the problem of representing physical observable particles.
Among the operators (\ref{A}) we distinguish some that change the number of
particles by one: 
\begin{equation}
b_{m}\left( k\right) =\widehat{A}_{m,m+1}\left( k\right)
,\;\;b_{m}^{+}\left( k\right) =\widehat{A}_{m+1,m}\left( k\right) ,
\label{b}
\end{equation}
\begin{equation}
\left[ \widehat{n},b_{m}^{(+)}\left( k\right) \right] =\pm b_{m}^{(+)}\left(
k\right) ,\;\;\;\left[ H_{0},b_{m}^{(+)}\left( k\right) \right] =\pm \omega
_{k}b_{m}^{(+)}\left( k\right) ,
\end{equation}
where 
\begin{equation}
\widehat{n}=\sum_{k}\widehat{n}_{k}=\sum_{n,k}nN_{n,k}.  \label{Nk}
\end{equation}
Then we can define the ground state $\Phi _{0}$ (where $b_{m}\left( k\right)
\Phi _{0}=0$ for arbitrary $m=0,1,\ldots $ and $k$) which corresponds to the
minimum energy for a fixed mode distribution $N_{k}$. Note that in contrast
to the standard theory, the ground state is generally characterized by a
nonvanishing particle number density $\widehat{n}\Phi _{0}=n_{0}\Phi _{0}$
and is not a Lorentz invariant state.

In the absence of processes related to changes in the topology of space and
for a mode distribution of the form $N_{k}=1$ (there is only one mode for
each wave number $k$), the standard field theory is restored. Furthermore,
there is a rather general case in which the concept of an effective field
can be introduced to restore the standard picture.

Indeed, consider the case in which the interaction operator in (\ref{Ham})
is expressed solely in terms of the set of operators $b_{0}\left( k\right) $
and $b_{0}^{+}\left( k\right) $ and the operator $N_{k}=\sum_{n}C^{+}\left(
n,k\right) C\left( n,k\right) $ is an integral of motion (i.e., $\left[
N_{k},H\right] =0$). Then we can renormalize the set of operators $%
b_{0}\left( k\right) $ and $b_{0}^{+}\left( k\right) $ as follows 
\begin{equation}
d_{k}=N_{k}^{-1/2}b_{0}\left( k\right)
,\;\;\;d_{k}^{+}=N_{k}^{-1/2}b_{0}^{+}\left( k\right) ,  \label{norm}
\end{equation}
and restore the standard commutation relations (this follows from (\ref{3.3m}%
) and (\ref{A})) 
\begin{equation}
\left[ d_{k},d_{k^{\prime }}^{+}\right] =\delta _{kk^{\prime }}.
\end{equation}
Thus, the standard algebra is restored and the set of operators \{$%
d_{k},d_{k}^{+}$\} can be used to define an effective field $\widetilde{%
\varphi }$ whose quanta coincide with physical particles in the complete
(extended) theory. One may consider the effective field as a renormalized
field $\varphi $. The basis of the Fock space will consist of the vectors $%
\left\{ d_{k_{1}}^{+}d_{k_{2}}^{+}\ldots d_{k_{n}}^{+}\Phi _{0}\right\} $.
The ground state $\Phi _{0}$ is, in general, not a Lorentz invariant state,
for it contains a nonvanishing particle number and energy densities: $%
\widehat{n}_{k}\Phi _{0}=\overline{n}_{k}\Phi _{0}$ and $H_{0k}\Phi
_{0}=e_{k}\Phi _{0}$, where $e_{k}=\overline{n}_{k}\omega _{k}$ and $%
\widehat{n}_{k}$ is the operator defined in (\ref{Nk}). We stress that this
does not mean that the Lorentz invariance of the theory is broken. This
situation is completely analogous to the spontaneous break of symmetry in
gauge theories. The physical particles are excitations of the ground state
and for the number of particles operator we get $d_{k}^{+}d_{k}=\delta 
\widehat{n}_{k}=\widehat{n}_{k}-\overline{n}_{k}$. However, the properties
of the ground state $\Phi _{0}$ remain beyond the scope of the effective
field and should be determined in the complete theory.

\section{Ground state and possible effects}

Consider now properties of the ground state $\Phi _{0}$. First, we note that
there exists a trivial ground state which coincides with the vacuum state of
the ordinary quantum field theory (in which $N_{k}=1$ for all $k$). In this
case our theory reduces to the standard one. However, we recall that in the
very beginning of the evolution our Universe has passed the quantum stage.
During this period processes involving topology changes take place and this
should result in some excess of the field degrees of freedom (modes). After
the quantum stage, processes with topology changes are suppressed (e.g., see
experimental restrictions in Refs. \cite{osc}), and the structure of space
is preserved. Thus, we may expect that in the modern universe the ground
state has a nontrivial structure. In the simplest case the field ground
state $\Phi _{0}$ is characterized by occupation numbers of the type 
\begin{equation}
N_{k,n}=\theta \left( \mu -n\omega _{k}\right) ,  \label{GST}
\end{equation}
where $\theta \left( x\right) $ is the Heaviside step function and $\mu $ is
the chemical potential. Thus, for the mode spectral density we get 
\begin{equation}
N_{k}=\sum_{n=0}^{\infty }\theta \left( \mu -n\omega _{k}\right) =1+\left[ 
\frac{\mu }{\omega _{k}}\right] \,\,,  \label{3.9m}
\end{equation}
where $[x]$ denotes the integer part of the number $x$. Equation (\ref{3.9m}%
) shows that at $\omega _{k}>\mu $ we have $N_{k}=1$, i.e., the field
structure corresponds to a flat Minkowski space, while in the range $\omega
_{k}<\mu $ we have $N_{k}>1$ and the field should exhibit a nontrivial
properties. In particular, this state is characterized by the spectral
density of the number of particles 
\begin{equation}
\overline{n}_{k}=\sum_{n=0}^{\infty }n\theta \left( \mu -n\omega _{k}\right)
=\frac{1}{2}\left( 1+\left[ \frac{\mu }{\omega _{k}}\right] \right) \,\left[ 
\frac{\mu }{\omega _{k}}\right] \,
\end{equation}
and the spectral density of the ground-state energy 
\begin{equation}
e_{k}=\omega _{k}\overline{n}_{k}=\frac{\omega _{k}}{2}\left( 1+\left[ \frac{%
\mu }{\omega _{k}}\right] \right) \,\left[ \frac{\mu }{\omega _{k}}\right] .
\label{H0}
\end{equation}

Since the given particles correspond to the ground state of the field, in
ordinary processes (which do not involve topology changes) these particles
are not manifested explicitly. From the effective-field standpoint, such
particles are ''dark'' and should comprise dark matter of the universe
(properties of this dark matter were considered in Ref.\cite{k99}). In what
follows we discuss one main effect which admit experimental verification. It
is an increase in the level of fluctuations in field potentials. In the case
of massless particles field potentials are measurable quantities and are of
special interest.

Consider fluctuations of the field potentials in the ground state (\ref{GST}%
). From (\ref{field}), (\ref{A}), and (\ref{GST}) we find 
\begin{equation}
\left\langle \varphi \left( x\right) \varphi \left( x+r\right) \right\rangle
=\frac{1}{\left( 2\pi \right) ^{2}}\int\limits_{0}^{\infty }\frac{dk}{\omega
_{k}}\frac{\sin kr}{kr}\Phi ^{2}\left( k\right) ,  \label{poten}
\end{equation}
where 
\[
\Phi ^{2}\left( k\right) =k^{2}N_{k}=k^{2}\left( 1+\left[ \frac{\mu }{\omega
_{k}}\right] \right) . 
\]
Thus, the presence of the ''dark'' particles in the ground state provides
the increase in the level of quantum fluctuations of field potentials $%
\delta \Phi ^{2}=k^{2}\left[ \frac{\mu }{\omega _{k}}\right] $ in comparison
with the standard vacuum noise ($\mu =0$). In the case of massless fields
this increase is substantial at long wavelengths $k\ll \mu $, where $\delta
\Phi ^{2}\simeq k\mu $.

In the present universe fields are characterized by a nonvanishing
temperature $T$. Thus, instead of the ground state we should use the thermal
equilibrium state which is characterized by occupation numbers of the type $%
N_{k,n}=\left( \exp \left( \frac{\mu -n\omega _{k}}{T}\right) +1\right) ^{-1}
$. $\ $Since processes associated with topology changes are the first to
stop during the early stages of the evolution of the universe, one should
expect that $T\ll T_{\gamma }$ ($T_{\gamma }$ is temperature of the
microwave background radiation). On the other hand, for the value of $\mu $
one can obtain an upper bound $\mu ^{\ast }\sim 10^{2}T_{\gamma }$ (which
corresponds to the situation when all observable dark matter is formed by
the hidden particles). Then for massless fields we find that at long
wavelengths $k\ll T\ll \mu $ field fluctuations turn out to be
scale-independent $\delta \Phi ^{2}\simeq T\mu $. In the case of the
electromagnetic field such fluctuations can be easily observed. Indeed, in
such a field any conductive sample will exhibit current fluctuations with
spectrum $1/f$. We note that unlike the standard $1/f$-noise these
fluctuations have the external origin and, therefore, have a universal
nature. In particular they do exist in the absence of a constant current in
the sample.

In conclusion we consider a renormalization of masses of physical particles
which takes place in case of fields with self-action and caused by the
presence of the hidden particles in the ground state. Consider as an example
the real scalar field with the potential in (\ref{Hf}) of the type $V=\frac{%
\lambda }{4!}\varphi ^{4}$. Then the observable value of the mass of
particles will be $m_{ph}^{2}=m_{0}^{2}+\frac{\lambda }{2}\left\langle
\varphi ^{2}\right\rangle $, where $m_{0}^{2}$ is the initial value (for $%
\mu =0$) and we assume $\lambda \ll 1$. Thus we find 
\begin{equation}
\delta m^{2}=\left\{ 
\begin{array}{c}
\frac{\lambda }{\left( 4\pi \right) ^{2}}\xi \left( 2\right) \mu
^{2}\;\;as\;\;m_{0}=0, \\ 
\frac{\lambda }{24\pi ^{2}}\frac{z^{3}}{m_{0}}\;\;\;as\;\;m_{0}\neq 0,
\end{array}
\right.
\end{equation}
where $z^{2}=\mu ^{2}-m_{0}^{2}$, $\xi \left( 2\right) =\sum_{n=1}^{\infty }%
\frac{1}{n^{2}}$, and we assume $z\ll m_{0}$.

\section{Summary}

In this manner the extension of the field theory proposed predicts a number
of new effects which in principle admit experimental verification. First of
all the ground state of the theory, which represents the vacuum state from
the standard field theory standpoint, is characterized by a finite density
of particles. These particles are dark and contribute to the dark matter of
the universe. The presence of particles in the ground state results in the
fact that this state is not a Lorentz invariant state, though the complete
theory remains to be invariant one. This situation is similar to the
spontaneous symmetry breaking in gauge theories. 

The presence of the dark particles leads also to an additional
renormalization of parameters of physical particles and an increase in the
level of quantum fluctuations in field potentials. The last effect is
crucial for the experimental verification of the theory proposed. Indeed,
for massless fields the increase is essential at small wavelengths and in
the presence of a temperature field fluctuations are scale-independent. This
means that the present universe is filled with a random electromagnetic
field which at scales $k\ll T_{\gamma }$ ($T_{\gamma }$ is the temperature
of the microwave background radiation) has $1/f$ spectrum $\delta
E^{2}\simeq T_{\gamma }\mu \leq 10^{2}T_{\gamma }^{2}$. In any electric
circuit this field should cause an additional additive $1/f$-noise, which
due to the external origin should be correlated in different circuits.
Direct observations of such a noise will allow to determine the value or
limits of the parameter $\mu $ which characterizes the nontrivial
topological structure of space.

\bigskip

This work was supported by grants from Russian Fund for Fundamental Research
(Grant No. 98-02-16273) and DFG (Grant No. 436 RUS 113/236/0(R)).

\end{document}